\documentclass[aps,preprint,amsmath,amssymb]{revtex4-1}
\usepackage{slashed}
\usepackage{graphicx}
\usepackage{color}
\newcommand{\nn}{\nonumber}
\newcommand{\bd}{\begin{document}}
\newcommand{\ed}{\end{document}}
\newcommand{\bc}{\begin{center}}
\newcommand{\ec}{\end{center}}
\newcommand{\be}{\begin{eqnarray}}
\newcommand{\ee}{\end{eqnarray}}
\newcommand{\ba}{\begin{array}}
\newcommand{\ea}{\ed{array}}
\newcommand{\strich}[1]{#1  \! \! \slash}
\newcommand{\eqn}{\global\def\theequation}
\newcommand{\sw}{sin^2 \theta_W}
\newcommand{\fbd}{f_B}
\renewcommand{\thefootnote}{\alph{footnote}}
\newcommand{\se}{\section}
\newcommand{\sse}{\subsection}
\newcommand{\bi}{\bibitem}
\def\figcap{\section*{Figure Captions\markboth
     {FIGURECAPTIONS}{FIGURECAPTIONS}}\list
     {Figure \arabic{enumi}:\hfill}{\settowidth\labelwidth{Figure 999:}
     \leftmargin\labelwidth
     \advance\leftmargin\labelsep\usecounter{enumi}}}
\let\endfigcap\endlist \relax
\def\reflist{\section*{References\markboth
     {REFLIST}{REFLIST}}\list
     {[\arabic{enumi}]\hfill}{\settowidth\labelwidth{[999]}
     \leftmargin\labelwidth
     \advance\leftmargin\labelsep\usecounter{enumi}}}
\let\endreflist\endlist \relax

\def\Journal#1#2#3#4{{#1} {{\bf #2},} {#4} {(#3)}}
\def\NCA{Nuovo Cimento}
\def\NIM{Nucl. Instrum. Methods}
\def\NIMA{{Nucl. Instrum. Methods} A}
\def\NP{{Nucl. Phys.} }
\def\NPB{{Nucl. Phys.} B }
\def\NPA{{Nucl. Phys. A}}
\def\PLB{{Phys. Lett.}  B}
\def\PL{{Phys. Lett.}}
\def\PPSA{{Proc. Phys. Soc.} A}
\def\PRP{{ Phys. Rep.}}
\def\PRL{ Phys. Rev. Lett.}
\def\PR{{Phys. Rev.}}
\def\PRD{{Phys. Rev.} D}
\def\PRC{{Phys. Rev.} C}
\def\ZP{{Z. Phys.}}
\def\ZPC{{Z. Phys. C}}
\def\EPJ{{Eur. Phys. J.}}
\def\EPJC{{Eur. Phys. J.} C}
\def\ZPA{{Z. Phys.} A}
\def\MPL{{Mod. Phys. Lett.}}
\def\MPLA{{Mod. Phys. Lett.} A}
\def\CPC{Comput. Phys. Commun.}
\def\JHEP{{J. High Energy Phys.}}
\def\JPG{{J. Phys. G.}}
\def\SJNP{Sov. J. Nucl. Phys.}
\def\NCA{ Nuovo Cimento}
\def\NIM{ Nucl. Instrum. Methods}
\def\NIMA{{ Nucl. Instrum. Methods} A}
\def\NP{{ Nucl. Phys.}}
\def\ANP{{Adv. Nucl. Phys.}}
\def\CPC{{Comput. Phys. Commun.}}

\begin{document}
\title{Weak Decays of the $\Lambda_{b}$ Baryon in Light-Front Dynamics}

\author{Chong-Chung Lih$^{1,2}$\footnote{cclih123@gmail.com}
, Cong-Yu Li$^{2,3,4}$\footnote{licongyu24@mails.ucas.ac.cn}
and Chao-Qiang Geng$^{1,2}$\footnote{cqgeng@ucas.ac.cn}
}

\affiliation{
$^{1}$Synergetic Innovation Center for Quantum Effects and Applications,
Hunan Normal University, Changsha 410081, China\\
$^{2}$School of Fundamental Physics and Mathematical Sciences,
Hangzhou Institute for Advanced Study, UCAS, Hangzhou 310024, China\\
$^{3}$Institute of Theoretical Physics,~UCAS, Beijing 100190, China\\
$^{4}$University of Chinese Academy of Sciences, 100190 Beijing, China
}

\date{\today}

\begin{abstract}

We investigate the exclusive semileptonic and nonleptonic 
$\Lambda_{b} \to \Lambda_{c}(p)$ decays within the Standard Model
by using the light-front quark model. 
To determine the behavior of the $\Lambda_{b} \to \Lambda_{c}(p)$ transition form factors, 
we employ the Bethe-Salpeter formalism in the timelike region, 
effectively accounting for both valence and nonvalence contributions.
Using these form factors, including nonvalence contributions, 
we obtain branching fractions of 
$\Lambda_b \to \Lambda_c\,\ell\,\bar{\nu}_{\ell}$, 
$\Lambda_b \to \Lambda_c\,\tau\,\bar{\nu}_{\tau}$, 
$\Lambda_b \to p\,\ell\,\bar{\nu}_{\ell}$
and 
$\Lambda_b \to p\,\tau\,\bar{\nu}_{\tau}$~($\ell=e$ or $\mu$)
are found to be around $5.39\%$, $1.41\%$, $3.33\times 10^{-4}$ and $2.08\times 10^{-4}$, respectively,  
which are consistent with the experimental measurements. 
The ratios of
$R^{\ell\tau}(\Lambda_{c})=\frac{{\cal B}({\Lambda_b}\to {\Lambda_c}\,\tau\,\bar{\nu}_\tau)}
{{\cal B}({\Lambda_b}\to {\Lambda_c}\,\ell\,\bar{\nu}_l)}$ and 
$R^{\ell\tau}(p)=\frac{{\cal B}({\Lambda_b}\to p\,\tau\,\bar{\nu}_\tau)}
{{\cal B}({\Lambda_b}\to p\,\ell\,\bar{\nu}_l)}$, 
are given by $0.261^{+0.097}_{-0.117}$ and $0.624^{+0.119}_{-0.129}$, 
respectively. The forward-backward asymmetries in the above semilepton decays are
also examined.
 Our results indicate that, 
within the light-front framework, the nonvalence contributions
to the  asymmetries are negligible 
compared to the $\beta$-induced uncertainties.  
In addition, we calculate the branching ratios for the nonleptonic decays  of
$\Lambda_{b}\to \Lambda_{c}(p)\,M$ with $M$ being the pseudo scalar and vector mesons
with  the results found to be consistent with the current experimental data. 
\end{abstract}
%\pacs{}

\maketitle

\se{Introduction}

Weak decays of bottom baryons provide an important testing ground
for the Standard Model and for our understanding of nonperturbative QCD dynamics.
Compared with mesons, baryons have a three-quark structure that 
leads to more complex nonperturbative dynamics 
and presents additional theoretical challenges. 
Among the weak decays of bottom baryons, semileptonic decays are of particular 
interest because they provide direct access to the 
Cabibbo-Kobayashi-Maskawa (CKM) matrix elements and allow for rigorous 
testing of lepton flavor universality (LFU). 
Furthermore, semileptonic observables involving different charged 
leptons have emerged as sensitive probes of possible physics beyond the Standard Model.
Nonleptonic decays of bottom baryons represent another important
area for studying flavor dynamics. These processes are crucial for 
understanding the mechanisms of hadronic weak decays, 
testing factorization hypotheses, and exploring CP violation.
Compared with mesonic decays, the three-quark structure of baryons leads
to more complex nonperturbative dynamics and presents additional theoretical challenges.

Reports from the PDG~\cite{PDG}, combined with recent experimental progress, 
have significantly enhanced our understanding of the weak decays of bottom baryons.
In particular, the DELPHI~\cite{DELPHI}, LHCb~\cite{LHCb1,LHCb2,LHCb3}, 
and CDF~\cite{CDF2,CDF3} collaborations have reported measurements of the branching fractions of 
semileptonic and nonleptonic $\Lambda_b$ decays. 
These measurements provide stringent tests of theoretical approaches 
and have motivated increasingly precise calculations of the underlying 
hadronic transition form factors. 
Although semileptonic and nonleptonic decays serve different phenomenological purposes, 
they share a common theoretical element: transition form factors describing the hadronic 
matrix elements between the initial and final baryon states. 
These form factors encode the long-distance QCD dynamics associated with 
weak transitions and are a major source of theoretical uncertainty. 
Considerable efforts have been devoted to determining these form factors 
using lattice QCD (LQCD)~\cite{LQCD1,LQCD2}, QCD sum rules~\cite{QCDSR1,QCDSR2,QCDSR3}, 
light-cone sum rules (LCSR)~\cite{LCSR,LCSR2,LCSRp}, 
perturbative QCD (PQCD)~\cite{PQCD,PQCD1,PQCD2,PQCD3,PQCD4}, 
and various other theoretical models~
\cite{RQM,HCQM,CCQM,CCQM1,CCQM2,BagModel,BSM,HQS221016825,QCDF,SM,LFQM1,LFQM,LFQM2,LFQM3,
Rahmani2025AFB}. 
Despite significant progress, noticeable discrepancies among different 
theoretical predictions remain, particularly in regions
where nonperturbative effects are substantial.

In this work, we employ the light-front quark model (LFQM), 
formulated within the framework of light-front quantization, 
to calculate hadronic transition form factors directly in the physical 
timelike region ($q^2>0$) and investigate weak decays of the $\Lambda_b$ 
baryon within the Standard Model (SM). 
The LFQM provides a relativistic and nonperturbative framework for 
describing hadronic bound states and calculating transition form factors. 
It allows a direct treatment of physical timelike processes and 
avoids a separate extrapolation from spacelike form factors. 
In this work, we adopt the diquark picture~\cite{diquark,WF1,diquark3,diquark4}, 
in which a baryon is described as a bound state of an active quark and a spectator diquark, 
reducing the original three-body problem to an effective two-body system. 
This approach preserves the fundamental dynamical properties of 
the system while maintaining computational feasibility.

However, in frames with $q^{+}\neq 0$, higher-Fock components of the hadron state 
can contribute to the transition matrix elements through nonvalence 
configurations~\cite{BS1,BS2,BS3,BS4}. 
Recently, 
an efficient method for handling the contributions of nonvalence quarks 
to baryons has been developed within a Bethe-Salpeter based 
light-front framework~\cite{lf1,lf2,lf3}. 
The resulting framework provides a systematic approach to 
improving the theoretical description of the weak decays of heavy hadrons. 
Applying this framework, we calculate the transition form factors relevant to 
$\Lambda_b\to\Lambda_c(p)$ transitions and employ them to 
study both semileptonic and two-body nonleptonic decays. 
Numerical predictions are presented for the branching fractions 
of 
$\Lambda_b$ semileptonic and two-body nonleptonic decay modes.  
This analysis results in a unified description of bottom-baryon weak decays within the 
light-front framework and offers useful theoretical inputs for 
ongoing and future experimental studies.

This paper is organized as follows. 
In Section II, we calculate the   
form factors for the baryonic transitions of $\Lambda_b\to \Lambda_c(p)$. 
Numerical results and discussions regarding the theoretical framework 
for the semileptonic and nonleptonic decays of  
$\Lambda_b\to \Lambda_c(p)\,\ell^- \bar{\nu}_\ell$, 
$\Lambda_{b}\to \Lambda_{c}(p)\,M$ and their leptonic forward-backward asymmetry 
are presented in Section III.
We conclude in Sec. IV.

\section{framework for transition form factors of $B_{i}\to B_{f}$}

In this section, we present the theoretical framework for the transition form factors
of $B_{i}\to B_{f}$ within the LFQM,  
where $B_i$ and $B_f$ denote the initial and final baryons, respectively.
The hadronic matrix element is expressed in terms of six 
invariant form factors $f(g)_{j}(q^{2})$ ($j = 1, 2, 3$), as
\be
&&\langle B_{f}(P^{\prime},S^{\prime}=
\frac{1}{2},S_{z}^{\prime})|\mathcal{\bar{Q}}\gamma^{\mu}
(1-\gamma_{5})b|B_{i}(P,S=\frac{1}{2},S_{z})\rangle \nonumber\\
&& =  \bar{u}(P^{\prime},S_{z}^{\prime})
\Big[\gamma^{\mu} f_{1}(q^{2})
+i \frac{f_{2}(q^{2})}{M}\sigma^{\mu\nu}q_{\nu}
+\frac{f_{3}(q^{2})}{M} q^{\mu}\Big]
u(P,S_{z})\nonumber\\
&&\quad-\bar{u}(P^{\prime},S_{z}^{\prime})
\Big[\gamma^{\mu} g_{1}(q^{2})
+i \frac{g_{2}(q^{2})}{M}\sigma^{\mu\nu}q_{\nu}
+\frac{g_{3}(q^{2})}{M} q^{\mu}\Big]\gamma_{5}u(P,S_{z}),
\label{transitionVA}
\ee
where $\mathcal{Q}$ denotes the quark field in the $\Lambda_b\to p( \Lambda_{c})$ transition, 
$q=P-P^{\prime}$, $M^{(\prime)}$, 
$P^{(\prime)}$ and $S_{z}^{(\prime)}$
represent the mass, momentum and spin of $B_{i(f)}$, respectively.
The transition matrix elements in Eq.~(\ref{transitionVA}) can be applied to
the helicity amplitudes,
given by~\cite{lf11,HA1,HA2}
\begin{eqnarray}
H^{V(A)}_{\lambda_{B_{f}} \lambda_{W}}
\equiv\langle B_{f}|(\mathcal{\bar{Q}} b)_{V(A)}|B_{i}\rangle
\varepsilon^{*\mu}_W\,,
\label{helicityA}
\end{eqnarray}
where $\varepsilon^\mu_W$ is the polarization of the W boson,
and the values $\lambda_{B_{f}}=\pm 1/2$
represent the helicity states of the final baryon.
The initial baryon helicity $\lambda_{B_{i}}$ is determined by the relationship 
$\lambda_{B_{i}}=\lambda_{B_{f}}-\lambda_{W}$, where $\lambda_{B_{f}}=\pm1/2$ and 
$\lambda_{W}=t, 0, \pm1$.

The helicity amplitudes are related to the form factors through the
following expressions:
\begin{align}
H_{\frac{1}{2},t}^{V} & =\frac{\sqrt{Q_{+}}}{\sqrt{q^{2}}}
\left((M-M^{\prime})f_{1}+\frac{q^{2}}{M}f_{3}\right),\nonumber \\
H_{\frac{1}{2},0}^{V} & =\frac{\sqrt{Q_{-}}}{\sqrt{q^{2}}}
\left((M+M^{\prime})f_{1}-\frac{q^{2}}{M}f_{2}\right),\nonumber \\
H_{\frac{1}{2},1}^{V} & =\sqrt{2Q_{-}}\left(-f_{1}
+\frac{M+M^{\prime}}{M}f_{2}\right),\nonumber \\
H_{\frac{1}{2},t}^{A} & =\frac{\sqrt{Q_{-}}}{\sqrt{q^{2}}}
\left((M+M^{\prime})g_{1}-\frac{q^{2}}{M}g_{3}\right),\nonumber \\
H_{\frac{1}{2},0}^{A} & =\frac{\sqrt{Q_{+}}}{\sqrt{q^{2}}}
\left((M-M^{\prime})g_{1}+\frac{q^{2}}{M}g_{2}\right),\nonumber \\
H_{\frac{1}{2},1}^{A} & =\sqrt{2Q_{+}}
\left(-g_{1}-\frac{M-M^{\prime}}{M}g_{2}\right).
\end{align}
where the subscript ``$t$" denotes $H_{\frac{1}{2},t}^{V(A)}$ arising 
from the temporal component of the current $(\bar q b)_{V(A)}$,
$q^{2}$ is the lepton pair invariant mass, and
$Q_{\pm}=(M \pm M^{\prime})^{2}-q^{2}$.
The negative helicity amplitudes are defined by
\begin{equation}
H_{-\lambda_{B_{f}},-\lambda_{W}}^{V}=H_{\lambda_{B_{f}},
\lambda_{W}}^{V}\quad\text{and}\quad H_{-\lambda_{B_{f}},-\lambda_{W}}^{A}
=-H_{\lambda_{B_{f}},\lambda_{W}}^{A},
\end{equation}
while the helicity amplitudes for the left-handed current are given by 
\begin{equation}
H_{\lambda_{B_{f}},\lambda_{W}}=H_{\lambda_{B_{f}},
\lambda_{W}}^{V}-H_{\lambda_{B_{f}},\lambda_{W}}^{A}.
\end{equation}
In order to calculate the form factors in the LFQM,
we treat the baryon as a bound state described 
in the quark-diquark picture of $q_1$ and $q_{2,3}$,
where $q_{2,3}$ are combined into a single diquark, expressed as $q_{[2,3]}$.
Explicitly, the baryon bound state with the total momentum
$P$ and spin $S$ can be written as~\cite{lf1,WF1,WF2}
\be
|B (P,S,S_{z})\rangle & = & \int\{d^{3}p_{1}\}
\{d^{3}p_{2}\}2(2\pi)^{3}\delta^{3}(P-p_{1}-p_{2})\nonumber \\
&  & \times\sum_{\lambda_{1},\lambda_{2}}\Psi^{SS_{z}}
(p_{1},p_{2},\lambda_{1},\lambda_{2})|q_{1}(p_{1},\lambda_{1})[q_2, q_3]
(p_{2},\lambda_{2})\rangle\,,
\label{boundstate}
\ee
where $q_{1}=b,~c$ and $u$ denotes the active quark corresponding to
$\Lambda_{b}$, $\Lambda_{c}$ and $p$, $[q_2, q_3]$ represents the diquark,
$\Psi^{SS_{z}}$ denotes the total wave function and 
$p_{1,2}$  are the on-shell LF momenta, which satisfy
\be
        && p^+_1=x_1 P^+, \quad p^+_2=x_2 P^+, \quad x_1+x_2=1\,,\nn \\
        && p_{1\bot}=x_1 P_\bot+k_\bot, \quad p_{2\bot}=x_2
        P_\bot-k_\bot\,,
\label{Pfraction}
\ee
with $(x_{1,2},k_\perp)$ being the longitudinal momentum fractions
and transverse momenta of the constituent particles.
Using the Melosh transformation~\cite{Melosh:1974cu},
we write the wave function in the following convenient form: 
\be
\Psi^{SS_{z}}(p_{1},p_{2},\lambda_{1},\lambda_{2})=
\frac{1}{\sqrt{2(p_{1}\cdot P+m_{1}M_{0})}}\bar{u}(p_{1},\lambda_{1})
\Gamma_{l,m} u(P,S_{z})\phi(x,k_{\perp})\,,
\label{1/2}
\ee
where $\Gamma_{l,m}$ is the coupling vertex function of the decaying quark $q_{1}$
and the diquark in the baryon state.
For the scalar diquark, the coupling vertex is $\Gamma_{l,m}=1$.
Regarding the distribution amplitude function of $\phi(x,k_{\perp})$ in Eq.~(\ref{1/2}), 
we use the following Gaussian-type function:

\be
\phi(x,k_{\perp})=4\left(\frac{\pi}{\beta^{2}}\right)^{3/4}\sqrt{\frac{dk_{z}}{dx}}\exp
\left(\frac{-\vec{k}^{2}}{2\beta^{2}}\right)\,,
\label{DA}
\ee
where $\beta$ is the baryon shape parameter and $k_z$ is defined by
\be
k_z=\frac{xM_0}{2}-\frac{m^2_{2}+k^2_{\perp}}{2xM_0}\,,~~~~~
M_0^2&=&{ m_{1}^2+k_\bot^2\over 1-x}+{ m_{2}^2+k_\bot^2\over  x}\,.
\ee
Using the bound states of
$|B_{i}(P,S,S_z)\rangle$ and $|B_{f}(P^{\prime},S^{\prime},S_z^{\prime})\rangle$
in Eq.~(\ref{boundstate}) and the above identities,
we derive the matrix elements of the baryonic transition
in the LF frame. For the $\mu=+$ component,
the transition matrix elements are given by
\be
&&\langle B_{f}(P^{\prime},S^{\prime},S_z^{\prime})|
\mathcal{\bar{Q}}\gamma^{+}(1-\gamma_{5}) b|B_{i}(P,S,S_z)\rangle\nonumber \\
&=&
N_{fs}\int{\{d^{4}p_{2}\}}
\frac{\chi_{B_{i}}(x,{\bf k}_{\perp}) I^{+}
\chi^{\prime}_{B_{f}}(x',{\bf k'}_{\perp})}{(p_{1}^{2}-m_{1}^{2}+i\epsilon)
(p_{1}^{'2}-m_{1}^{'2}+i\epsilon)}\,,
\label{matrix}
\ee
where $I^{+}=\sum_{\lambda_{2}}\bar{u}(P',S'_{z})
\left[\bar{\Gamma}^{\prime}_{S(A)}(\strich p_{1}^{\prime}+m_{1}^{\prime})
\gamma^{+}(1-\gamma_{5})(\strich p_{1}+m_{1})\Gamma_{S(A)}\right]u(P,S_{z})$,
$\bar \Gamma=\gamma^0 \Gamma^\dagger\gamma^0$,
$\chi_{B_{i}}$($\chi^{\prime}_{B_{f}}$) corresponds to the vertex function of the
baryon $B_{i(f)}$, and
the flavor spin factor $N_{fs}$ is a process dependent overlap factor for the particle transition,
given by the specific process.
Following Refs.~\cite{LFcal1,LFcal2,LFcal3,LFcal4}
we consider the case with ``$\mu = +$" and $q^{+} \neq 0$ 
and derive the following relations:
\be
&&\langle B_{f}(P^{\prime},S^{\prime},S_z^{\prime})|
V^{+}|B_{i}(P,S,S_z)\rangle
=-2\sqrt{P^{+}P^{\prime}}\bigg\{f_{1}(q^{2})\delta_{S^{\prime}_{z}S_{z}}
+\frac{f_{2} (q^{2})}{M}(\sigma \cdot q_{\perp})\sigma^{3}\,\delta_{S^{\prime}_{z}S_{z}}
\nonumber \\
&&+\frac{f_{3}(q^{2})}{2M}
\left[\left(\frac{M}{P^{+}}+\frac{M^{\prime}}{P^{\prime +}}\right)
\delta_{S^{\prime}_{z}S_{z}}+\left(\frac{P^{\prime}_{\perp}}{P^{\prime +}}-
\frac{P_{\perp}}{P^{+}}\right)\sigma^{3}\delta_{S^{\prime}_{z},-S_{z}}
\right]q^{+}\bigg\}
\nonumber \\
&&\langle B_{f}(P^{\prime},S^{\prime},S_z^{\prime})|
A^{+}|B_{i}(P,S,S_z)\rangle
=2\sqrt{P^{+}P^{\prime}}\bigg\{g_{1}(q^{2})\sigma^{3}\delta_{S^{\prime}_{z}S_{z}}
+\frac{g_{2}(q^{2})}{M}(\sigma \cdot q_{\perp})\delta_{S^{\prime}_{z}S_{z}}
\nonumber \\
&&-\frac{g_{3}(q^{2})}{2M}
\left[\left(\frac{M}{P^{+}}-\frac{M^{\prime}}{P^{\prime +}}\right)
\sigma^{3}\delta_{S^{\prime}_{z}S_{z}}-\left(\frac{P^{\prime}_{\perp}}{P^{\prime +}}-
\frac{P_{\perp}}{P^{+}}\right)\delta_{S^{\prime}_{z},-S_{z}}
\right]q^{+}\bigg\}\,.
\label{LFff}
\ee
Using the orthogonality relation involving between $\delta_{S^{\prime}_{z}S_{z}}$,
we notice that $f_{1}(q^{2})(g_{1}(q^{2}))$ and $f_{3}(q^{2})(g_{3}(q^{2}))$
are related through a linear combination but are both orthogonal to $f_{2}(q^{2})(g_{2}(q^{2}))$. 
Since the system is analyzed in the time-like domain
(i.e., $q^2 = q^{+}q^{-} - q^2_{\perp} \geq 0$), the momentum is only
transmitted longitudinally, i.e., $q_{\perp} = 0$. 
Therefore, when calculating the matrix elements on the left-hand side of 
Eq.~($\ref{LFff}$), we initially take $q_{\perp} \not= 0$  
and set it to zero only after extracting the form factors. 
On the right-hand side of Eq.~(\ref{LFff}), we have 
$P(P^{\prime})_{\perp}=0$. 
Combined this result with Eq.~(\ref{matrix}), we express the form factors as follows:

\be
&&f_{1}(q^{2})\delta_{S^{\prime}_{z}S_{z}}
+\frac{f_{3}(q^{2})}{2M} \left(\frac{M}{P^{+}}
+\frac{M^{\prime}}{P^{\prime +}}\right)
q^{+}\,\delta_{S^{\prime}_{z}S_{z}}  \nonumber \\
&=& f_{1}(q^{2}) + A f_{3}(q^{2}) = Hv(q^{2}) \nonumber \\
&=& N_{fs}\int{\{d^{4}p_{2}\}}
\frac{\chi_{B_{i}}(x,{\bf k}_{\perp})
\chi^{\prime}_{B_{f}}(x',{\bf k'}_{\perp})}{(p_{1}^{2}-m_{1}^{2}+i\epsilon)
(p_{1}^{'2}-m_{1}^{'2}+i\epsilon)}\,\nonumber \\
&&\times {\rm Tr} [ (\slashed P+M_{0})\gamma^{+}(\slashed P^{\prime}+M_{0}^{\prime})
(\slashed p_{1}^{\prime}+m_{1}^{\prime})\gamma^{+}
(\slashed p_{1}+m_{1}) ]\,, \nonumber \\
&&g_{1}(q^{2})\sigma^{3}\delta_{S^{\prime}_{z}S_{z}}
-\frac{g_{3}(q^{2})}{2M} \left(\frac{M}{P^{+}}
-\frac{M^{\prime}}{P^{\prime +}}\right)
\,\sigma^{3}q^{+}\,\delta_{S^{\prime}_{z}S_{z}}\nonumber \\
&=& g_{1}(q^{2}) + B g_{3}(q^{2}) = Ha(q^{2})\nonumber \\
&=& N_{fs}\int{\{d^{4}p_{2}\}}
\frac{\chi_{B_{i}}(x,{\bf k}_{\perp})
\chi^{\prime}_{B_{f}}(x',{\bf k'}_{\perp})}{(p_{1}^{2}-m_{1}^{2}+i\epsilon)
(p_{1}^{'2}-m_{1}^{'2}+i\epsilon)} \,\nonumber \\
&&\times {\rm Tr}[ (\slashed P+M_{0})\gamma^{+}\gamma_{5}(\slashed P^{\prime}+M_{0}^{\prime})
(\slashed p_{1}^{\prime}+m_{1}^{\prime})\gamma^{+}\gamma_{5}
(\slashed p_{1}+m_{1}) ]\,.
\label{f1g3}
\ee
\be
&&\frac{f_{2}(q^{2})}{M}(\sigma\cdot q_{\perp}\sigma^{3})\delta_{S^{\prime}_{z}S_{z}}\nonumber \\
&=& N_{fs}\int{\{d^{4}p_{2}\}}
\frac{\chi_{B_{i}}(x,{\bf k}_{\perp})
\chi^{\prime}_{B_{f}}(x',{\bf k'}_{\perp})}{(p_{1}^{2}-m_{1}^{2}+i\epsilon)
(p_{1}^{'2}-m_{1}^{'2}+i\epsilon)}\,\nonumber \\
&&\times{\rm Tr} [ (\slashed P+M_{0})\sigma^{\nu +}(\slashed P^{\prime}+M_{0}^{\prime})
(\slashed p_{1}^{\prime}+m_{1}^{\prime})\gamma^{+}
(\slashed p_{1}+m_{1}) ]\,\,, \nonumber \\
&&\frac{g_{2}(q^{2})}{M}(\sigma \cdot q_{\perp})\delta_{S^{\prime}_{z}S_{z}}
\nonumber \\
&=& N_{fs}\int{\{d^{4}p_{2}\}}
\frac{\chi_{B_{i}}(x,{\bf k}_{\perp})
\chi^{\prime}_{B_{f}}(x',{\bf k'}_{\perp})}{(p_{1}^{2}-m_{1}^{2}+i\epsilon)
(p_{1}^{'2}-m_{1}^{'2}+i\epsilon)}\,\nonumber \\
&&\times{\rm Tr} [ (\slashed P+M_{0})\sigma^{\nu +}\gamma_{5}(\slashed P^{\prime}+M_{0}^{\prime})
(\slashed p_{1}^{\prime}+m_{1}^{\prime})\gamma^{+}\gamma_{5}
(\slashed p_{1}+m_{1}) ]\,\,.
\label{f2g2}
\ee
where $\nu = 1, 2$ and $(x',{\bf k'}_\perp)$ are the LF relative momentum variables of
${\cal B}_{f}(P^{\prime},S^{\prime},S_z^{\prime})$
with the definitions given by replacing $x\to x'$ and ${\bf k}_\perp \to {\bf k'}_\perp$ 
in Eq.~(\ref{Pfraction}). 
In Eq.~(\ref{f1g3}), the variables $A$ and $B$ are written as follows:

\be
A= \frac{1}{2M}\left(\frac{M}{P^{+}}
+\frac{M^{\prime}}{P^{\prime +}}\right)q^{+}
=\frac{1}{2M}\frac{1-\alpha}{\alpha}(\alpha M+M^{\prime})\nonumber \\
B= -\frac{1}{2M}\left(\frac{M}{P^{+}}
-\frac{M^{\prime}}{P^{\prime +}}\right)q^{+}
=-\frac{1}{2M}\frac{1-\alpha}{\alpha}(\alpha M-M^{\prime})
\label{AB}
\ee
where the minus sign in the axial contribution has been absorbed into the definition of $B$
and $\alpha=P^{\prime+} /P^{+}$. There are two solutions for $\alpha$, namely:
\be
\alpha_{\pm}
=\frac{M^{2}+M^{\prime 2}-q^{2}\pm\sqrt{(M^{2}+M^{\prime 2}-q^{2})^{2}
-4 M^{2} M^{\prime 2}} }{2 M^{2}}\,.
\label{eq21}
\ee
where the $+(-)$ sign corresponds to the outgoing baryon recoiling
in the positive (negative) z-direction relative to the incoming baryon. 
The form factors are independent of the choice of 
reference frame along the direction of motion.
Therefore, for $f_{1}$ and $f_{3}$, we can derive
\be
f_{1}(q^2)&=&\frac{A_{-} Hv(q^{2})|_{\alpha=\alpha_{+}}
-A_{+} Hv(q^{2})|_{\alpha=\alpha_{-}}}{A_{-}-A_{+}},\nonumber \\
f_{3}(q^2)&=&\frac{Hv(q^{2})|_{\alpha=\alpha_{-}}
-Hv(q^{2})|_{\alpha=\alpha_{+}}}{A_{-}-A_{+}}.
\ee
where $A_{+}=A|_{\alpha=\alpha_{+}}$ and $A_{-}=A|_{\alpha=\alpha_{-}}$. 
Similarly, the expressions for $g_{1}$ and $g_{3}$ are 
obtained by making the replacements $A \to B$ and $H_{v}\to H_{a}$.

\begin{figure}[h]
\includegraphics{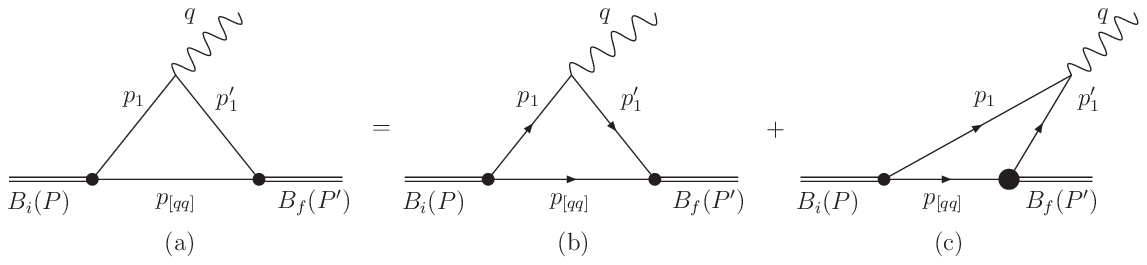}
\vskip 4cm
\caption{The effective treatment of the LF amplitude (a) can be decomposed into
the LF valence part (b) in $0 < x <\alpha$ and the nonvalence one (c)
in $\alpha < x < 1$, where
the small and large solid dots of the mediator-quark vertices
in (b) and (c) represent the LF ordinary
and nonvalence wavefunction vertices, respectively.}
\label{fig1}
\end{figure}
\begin{figure}[h]
\centering
\includegraphics{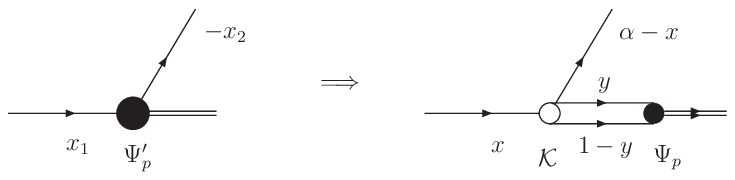}
\vskip 4cm
\caption{Relationship between the ordinary LF
wave function (small solid dot) and the nonvalence vertex (large solid dot)}
\label{figki}
\end{figure}

As shown in Fig.~\ref{fig1}, the full matrix element can be decomposed into contributions 
from the valence and nonvalence regions. 
Within the valence region, one has that $0 < x < \alpha$, $0 < p_{[qq]}^{+} < P'^{+}$
and $p^{-}_{[qq]} = p^{-}_{[qq]on}$, where the subscript ``$on$"
indicates the on-shell condition, as shown in Fig.~1b.
Fig.~1c shows the condition within the non-valence region,
where $\alpha < x < 1$, $P'^{+} < p_{[qq]}^{+} < P^{+}$
and $p_{1}^{-} = p^{-}_{1on}$.
The trace term in Eq.~(\ref{matrix}) can be written as the sum of the
valence $I^{+}_{V}$ and nonvalence $I^{+}_{NV}$ contributions.
Referring to Refs.~\cite{LFcal1,LFcal2,LFcal3,LFcal4},
we derive the transition form factors of the valence region
from Eqs.~(\ref{f1g3}) and (\ref{f2g2}) as follows:
\be
Hv_{val}(q^{2})&=&\frac{N_{fs}}{16\pi^3}
\int^{\alpha}_{0}dx\int d^{2}{\bf k}_{\perp}
\Phi_{B_{i}}(x,{\bf k}_{\perp}) \Phi_{B_{f}}(x',{\bf k'}_{\perp})
\nonumber \\
&&\times[k_{\perp}\cdot k_{\perp}^{\prime}+((1-x) M_{0}+m_{1})
((1-x^{\prime})M_{0}^{\prime}+m_{1}^{\prime})]\nonumber \\
Ha_{val}(q^{2})&=& \frac{N_{fs}}{16\pi^3}
\int^{\alpha}_{0}dx\int d^{2}{\bf k}_{\perp}
\Phi_{B_{i}}(x,{\bf k}_{\perp}) \Phi_{B_{f}}(x',{\bf k'}_{\perp})
\nonumber \\
&&\times[-k_{\perp}\cdot k_{\perp}^{\prime}+((1-x) M_{0}+m_{1})
((1-x^{\prime})M_{0}^{\prime}+m_{1}^{\prime})]\,.
\label{Hva}
\ee
\be
\frac{f_{2_{val}}(q^{2})}{M}&=& \frac{N_{fs}}{16\pi^3 q_{\perp}^{2}}
\int^{\alpha}_{0}dx\int d^{2}{\bf k}_{\perp}
\Phi_{B_{i}}(x,{\bf k}_{\perp})
\Phi_{B_{f}}(x',{\bf k'}_{\perp})\nonumber \\
&&\times[-(m_{1}+(1-x) M_{0})k_{\perp}^{\prime}\cdot q_{\perp}
+(m_{1}^{\prime}+(1-x^{\prime})M_{0}^{\prime})k_{\perp}\cdot q_{\perp}]\nonumber \\
\frac{g_{2_{val}}(q^{2})}{M}&=& \frac{N_{fs}}{16\pi^3 q_{\perp}^{2}}
\int^{\alpha}_{0}dx\int d^{2}{\bf k}_{\perp}
\Phi_{B_{i}}(x,{\bf k}_{\perp})
\Phi_{B_{f}}(x',{\bf k'}_{\perp})\nonumber \\
&&\times[-(m_{1}+(1-x) M_{0})k_{\perp}^{\prime}\cdot q_{\perp}
-(m_{1}^{\prime}+(1-x^{\prime})M_{0}^{\prime})k_{\perp}\cdot q_{\perp}]\,.
\label{f2g2val}
\ee
and
\be
\Phi_{B}(x,{\bf k}_{\perp})=\frac{\phi(x,k_{\perp})}
{\sqrt{2x_{1}(p_{1}\cdot P+m_{1}M_{0})}}
\ee
In Fig.~1c, the large solid black dot represents the nonvalence wave function vertex.
The vertex of the nonvalence wavefunction can usually be obtained from
the Bethe-Salpeter (B-S) theory by expanding the normal
B-S amplitude ~\cite{FVNV,BS1}.
The corresponding light-front bound-state equation can be written as~\cite{BS1,BSEQ2,BSEQ3}
\be
(M^{2}-M^{2}_{0})\Phi'(x_{i},{k}_{i\perp})
=\int [dy][d^2{\bf l}_{\perp}]
{\cal K}(x_{i},{\bf k}_{i\perp};y,{\bf l}_{\perp})
\Phi(y,{\bf l}_{\perp})\,.
\label{LFBS}
\ee
Both valence and nonvalence B-S amplitudes can be regarded as solutions of Eq.~(\ref{LFBS}).
The normal and nonvalence B-S amplitudes correspond
to  $x < \alpha$ and  $x > \alpha$, respectively.
In Fig. 1c, the nonvalence B-S amplitude can be effectively related to the valence B-S 
amplitude through analytic continuation at $x = \alpha$. 
In the LFQM, the relationship between the B-S amplitudes in the 
two regions is given in Refs.~\cite{BS1,BS2,BS3}.
The relevant function ${\cal K}$ in Eq.~(\ref{LFBS}) is the B-S core,
which in principle contains contributions from higher Fock states.
As shown in Fig.~\ref{figki}, the kernel ${\cal K}$ provides the
dynamical connection between the higher-Fock and ordinary valence configurations. 
However, obtaining the kernel in the integral equation, Eq.~(\ref{LFBS}), 
requires a nonperturbative QCD method. 
Using the following identity, the trace term in Eq.~(\ref{matrix})
can be decomposed into $I^{\mu}_{on}$ and $I^{\mu}_{inst}$.
\be
\slashed p+m=(\slashed p_{on}+m)+\frac{1}{2}\gamma^{+}(p^{-}-p^{-}_{on})\,.
\label{eq24}
\ee
The effective contribution is given by
\be
{\cal M}_{non-val}&=&\frac{N_{fs}}{16\pi^3}
\int^{1}_{\alpha}dx
\int d^{2}{\bf k}_{\perp}\Gamma_g(x,{\bf k}_{\perp})I^{+}_{NV}
\Phi_{B_{i}}(x,{\bf k}_{\perp})\nonumber\\
&\times&
\int \frac{dy}{y(1-y)}\int d^2{\bf l}_{\perp}
{\cal K}(x,{\bf k}_{\perp};y,{\bf l}_{\perp})
\Phi_{B_{f}}(y,{\bf l}_{\perp})\,.
\label{NVmatrix}
\ee
where $I^{+}_{NV}$ is the trace term of the non-valence region.
Substituting Eq.~(\ref{eq24}) into Eqs.~(\ref{f1g3}) and (\ref{f2g2}),
we obtain the trace term
$I^{+}_{NV}=I^{+}_{on}(p_{i}^{-}=p^{-}_{ion}=(m_{i}^{2}+k^{2}_{i\perp})/p^{+}_{i})+I^{+}_{inst}$.
The LF vertex function of the gauge boson $\Gamma_g$ corresponds to the
LF energy denominator in Eq.~(\ref{NVmatrix}),
and its explicit form is given by \cite{BS1,BS3}.
\be
\Gamma_g^{-1}(x,{\bf k}_\perp)=
\alpha\biggl[\frac{q^2}{1-\alpha} -
\biggl(\frac{{\bf k}^2_\perp + m^2_1}{1-x}
+\frac{{\bf k'}^2_\perp + m'^{2}_{1}}{\alpha-x}\biggr)
\biggr].
\label{gaugeWF}
\ee
The form factors related to the nonvalence diagram $I^+_{NV}$ are given by
\be
Hv_{NV}(q^{2})&=&\frac{N_{fs}}{16\pi^3}
\int^{1}_{\alpha}dx
\int d^{2}{\bf k}_{\perp}\Gamma_g(x,{\bf k}_{\perp})\Phi_{B_{i}}(x,{\bf k}_{\perp}) \nonumber \\
&\times&\{[k_{\perp}\cdot k_{\perp}^{\prime}+((1-x)M_{0}+m_{1})
((1-x^{\prime})M_{B_{f}}+m_{1}^{\prime})]+I^{+}_{inst}\}\nonumber\\
&\times&
\int \frac{dy}{y(1-y)}\int d^2{\bf l}_{\perp}
{\cal K}(x,{\bf k}_{\perp};y,{\bf l}_{\perp})
\Phi_{B_{f}}(y,{\bf l}_{\perp})\,, \nonumber\\
Ha_{NV}(q^{2})&=& \frac{N_{fs}}{16\pi^3}
\int^{1}_{\alpha}dx
\int d^{2}{\bf k}_{\perp}\Gamma_g(x,{\bf k}_{\perp})\Phi_{B_{i}}(x,{\bf k}_{\perp})\nonumber\\
&\times&\{[-k_{\perp}\cdot k_{\perp}^{\prime}+((1-x)M_{0}+m_{1})
((1-x^{\prime})M_{B_{f}}+m_{1}^{\prime})]+I^{\prime +}_{inst}\}\nonumber\\
&\times&
\int \frac{dy}{y(1-y)}\int d^2{\bf l}_{\perp}
{\cal K}(x,{\bf k}_{\perp};y,{\bf l}_{\perp})
\Phi_{B_{f}}(y,{\bf l}_{\perp})\,,
\label{f1g1NV}
\ee
\be
\frac{f_{2_{NV}}(q^{2})}{M}&=& \frac{N_{fs}}{16\pi^3 q_{\perp}^{2}}
\int^{1}_{\alpha}dx
\int d^{2}{\bf k}_{\perp}\Gamma_g(x,{\bf k}_{\perp})\Phi_{B_{i}}(x,{\bf k}_{\perp})\nonumber\\
&\times&\{
[(m_{1}+(1-x)M_{0})k_{\perp}^{\prime}\cdot q_{\perp}
-(m_{1}^{\prime}+(1-x^{\prime})M_{B_{f}})k_{\perp}\cdot q_{\perp}]+I^{+}_{inst}\}
\nonumber\\
&\times&
\int \frac{dy}{y(1-y)}\int d^2{\bf l}_{\perp}
{\cal K}(x,{\bf k}_{\perp};y,{\bf l}_{\perp})
\Phi_{B_{f}}(y,{\bf l}_{\perp})\,,\nonumber\\
\frac{g_{2_{NV}}(q^{2})}{M}&=& \frac{N_{fs}}{16\pi^3 q_{\perp}^{2}}
\int^{1}_{\alpha}dx
\int d^{2}{\bf k}_{\perp}\Gamma_g(x,{\bf k}_{\perp})\Phi_{B_{i}}(x,{\bf k}_{\perp})\nonumber\\
&\times&\{
[-(m_{1}+(1-x)M_{0})k_{\perp}^{\prime}\cdot q_{\perp}
-(m_{1}^{\prime}+(1-x^{\prime})M_{B_{f}})k_{\perp}\cdot q_{\perp}]+I^{\prime +}_{inst}\}
\nonumber\\
&\times& \int \frac{dy}{y(1-y)}\int d^2{\bf l}_{\perp} 
{\cal K}(x,{\bf k}_{\perp};y,{\bf l}_{\perp}) \Phi_{B_{f}}(y,{\bf
l}_{\perp})\,, \label{f2g2NV} 
\ee 
where $I^{+}_{inst}=I^{\prime
+}_{inst}=0$ after taking the trace. The complete form factors are
$f(g)_{j}=f(g)_{j_{V}}+f(g)_{j_{NV}}$ ($j=1,2,3$). For
Eqs.~(\ref{f1g1NV}) and (\ref{f2g2NV}), we define
\be
G_{B_{i}B_{f}}\equiv\int[dy][d^2{\bf l}_{\perp}] 
{\cal K}(x_{i},{\bf k}_{i\perp};y,{\bf }_\perp)\Phi_{B_{f}}(y,{\bf
l}_{\perp})\,, 
\ee 
which depends only on $x$ and ${\bf k}_{\perp}$, where $[dy]=dy/(y(1-y))$. 
In this study, we approximate $G_{B_{i}B_{f}}$
as a constant because the incoming baryon wavefunction $\Phi_{B_{i}}(x,
k_{\perp})$ strongly suppresses the integrand in the nonvalence
contribution region, and the effective $x$ region is narrow. 
This makes the momentum dependence of $G_{B_{i}B_{f}}$ 
practically weak, and only within the region of the
small momentum transfer~\cite{BS1,BS2,BS3}. 
Under this suppression mechanism, $G_{B_{i}B_{f}}$ 
in the Gaussian-type baryon quark-diquark wavefunction can still 
be approximated as a constant.

\se{Numerical Results And Discussion}

\sse{Form factors}
To numerically evaluate the transition form factors in the LFQM, 
we choose to work directly in the timelike region. In our calculation, 
we take the limit as \(q_\perp \to 0\). The  input parameters~\cite{WF1} are given in Table~I,
\begin{table}[htbp]
\caption{Input parameters for the $\Lambda_{b} \to \Lambda_{c}(p)$ transitions (in GeV)}
\vskip 0.2in
\label{beta values}
\begin{tabular}{ c c c c c c c c c c c c c} \hline
$m_{u,d}$ & $~m_{c}$ & $~m_{b}$ & $~m_{[qq']}$\cite{diquark3,diquark4} & $~\beta_{b[qq]}$ 
& $~\beta_{c[qq]}$ & $~\beta_{u[qq]}$ 
& $~f_{\pi}$ & $~f_{K}$ & $~f_{D}$ & $~f_{D^{*}}$ & $~f_{D_{s}}$ & $~f_{D_{s}^{*}}$
\\ \hline \hline
$0.24$ & $1.3$ & $4.2$ & $1.25$ & $~0.66\pm0.04$ & $~0.54\pm0.03 $ & $~0.46\pm0.04$
 & $0.13$ & $~0.16$ & $~0.2$ & $~0.2$ & $~0.23$ & $~0.23$
\\ \hline
\end{tabular}
\end{table}
where the quark masses are the constituent masses used in the quark model.
In Eq.~(\ref{NVmatrix}), the factor \(G_{\mathcal B_i\mathcal B_f}\) 
is treated as a constant with a value in the range \(1.0 \sim 6.0\), 
which has been tested in some exclusive semileptonic mesonic decays 
and shown to be a good approximation for processes with 
small momentum transfers~\cite{BS1,BS2,BS3}. In our numerical 
evaluation, we take $G_{\Lambda_{b}\Lambda_{c}}=4$ and $G_{\Lambda_{b}p}=1$ , 
with the flavor-spin factors $N_{fs}(\Lambda_b \to \Lambda_c )=1$ 
and $N_{fs}(\Lambda_b \to p )=1/\sqrt{2}$~\cite{LFQM1}.

To describe the \(q^2\) dependences of the form factors, 
we adopt the double-pole parametrization:
\be
F(q^2)=\frac{F(0)}{1+a(q^2/M_{pole}^{2})+b(q^4/M_{pole}^{4})}\,,
\label{fit1}
\ee
where \(M_{\text{pole}}=M_{\Lambda_b}=5.62\,\text{GeV}\) 
is the $\Lambda_b$ mass scale, 
and the parameters \((F(0),a,b)\) are determined by numerical analysis. 
The results of this parametrization are listed in Table~\ref{Table1}.

In Table ~\ref{Table1}, all fit parameters are determined primarily from the 
momentum dependences of the form factors, 
rather than from the physical observables themselves. 
In previous LFQM literature, $f_3$ and $g_3$ are consistently neglected 
in the spacelike region because the condition $q^+ = 0$ is imposed. 
Physically, the effects of $f_3$ and $g_3$ are suppressed 
by lepton masses; however, their contributions are non-negligible 
for $\tau$ leptons or lighter hadrons.

\begin{table}[htbp]
\caption{Form factors of the ${\Lambda_b} \to {\Lambda_c}(p)$ transitions 
with valence+nonvalence contributions}
\vskip 0.2in
\label{Table1}
\begin{tabular}{|c||c|c|c|c|c|c|} \hline
${\Lambda_b} \to {\Lambda_c}$ & $f_{1}$ & $f_{2}$ & $f_{3}$ & 
$g_{1}$ & $g_{2}$ & $g_{3}$
\\ \hline \hline
$F(0)$ & $0.499^{+0.040}_{-0.045}$ & $-0.179^{+0.017}_{-0.015}$ & $-0.259^{+0.017}_{-0.014}$ 
& $0.486^{+0.038}_{-0.043}$ & $-0.024\pm0.005$ & $-0.613^{+0.040}_{-0.033}$
\\ \hline
$a$ & $-2.47^{+0.21}_{-0.25}$ & $-3.08^{+0.17}_{-0.21}$ & $-3.50^{+0.14}_{-0.17}$ 
& $-1.49^{+0.21}_{-0.26}$ & $-4.65^{+0.23}_{-0.25}$ & $-3.36^{+0.14}_{-0.17}$
\\ \hline
$b$ & $2.02^{+0.40}_{-0.31}$ & $4.09^{+0.30}_{-0.22}$ & $3.84^{+0.29}_{-0.23}$
& $3.21^{+0.11}_{-0.03}$ & $6.34^{+0.52}_{-0.46}$ & $3.72^{+0.27}_{-0.21}$
\\ \hline \hline
${\Lambda_b} \to p$ & $f_{1}$ & $f_{2}$ & $f_{3}$ & 
$g_{1}$ & $g_{2}$ & $g_{3}$
\\ \hline \hline
$F(0)$ & $0.137\pm0.033$ & $-0.073\pm0.017$ & $-0.042\pm0.0045$ 
& $0.132\pm0.031$ & $-0.023\pm0.007$ & $-0.04\pm0.006$
\\ \hline
$a$ & $-2.59\pm0.17$ & $-3.07\pm0.09$ & $-2.76\pm0.03$ 
& $-2.52^{+0.22}_{-0.24}$ & $-3.20^{+0.05}_{-0.04}$ & $-2.82\pm0.04$
\\ \hline
$b$ & $2.01^{+0.16}_{-0.15}$ & $2.69\pm0.050$ & $2.02\pm0.03$
& $3.60^{+0.09}_{-0.07}$ & $2.70^{+0.01}_{-0.03}$ & $2.24\pm0.02$
\\ \hline
\end{tabular}
\end{table}

\begin{figure}[t!]
\includegraphics[width=3.1in]{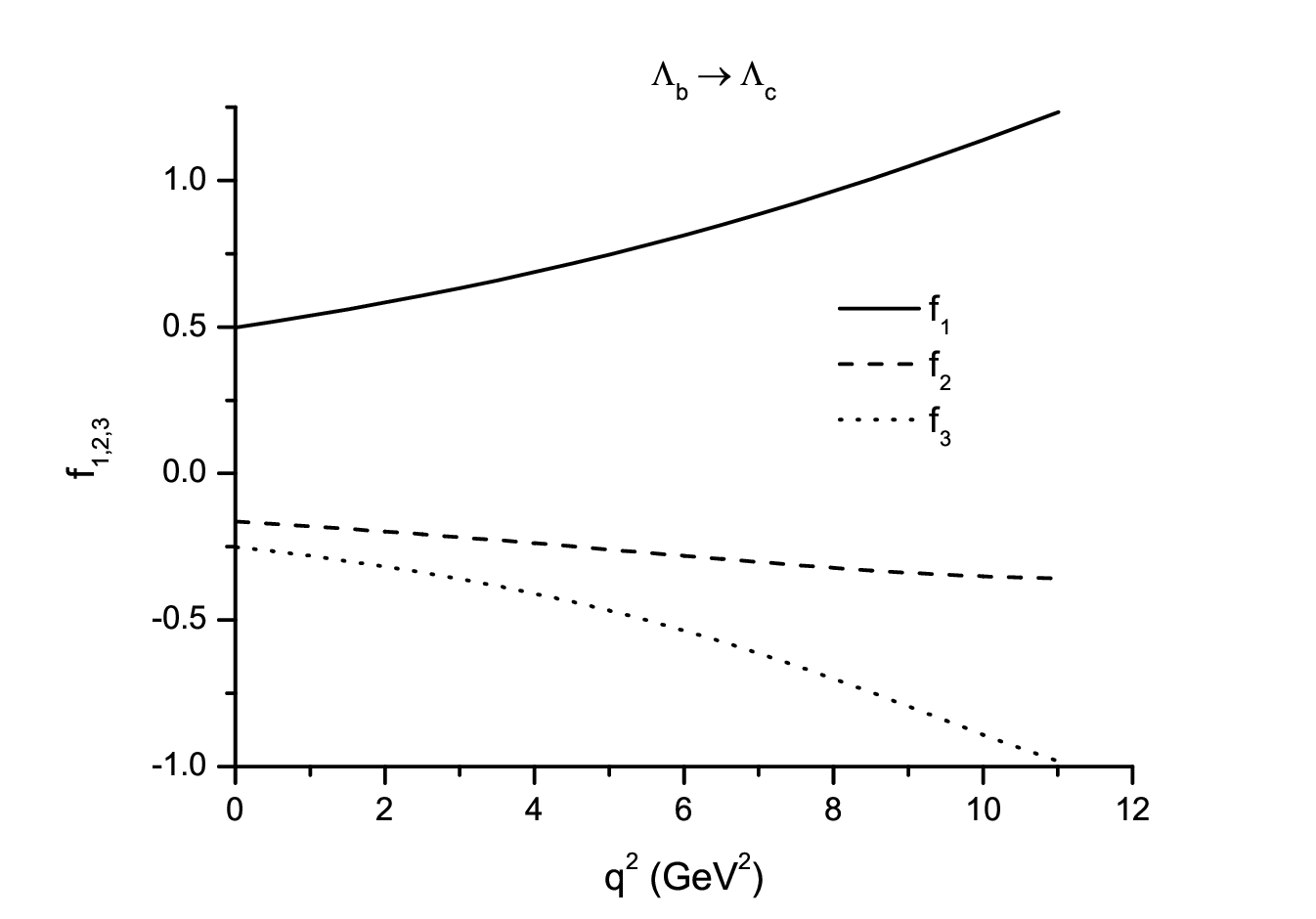}
\includegraphics[width=3.1in]{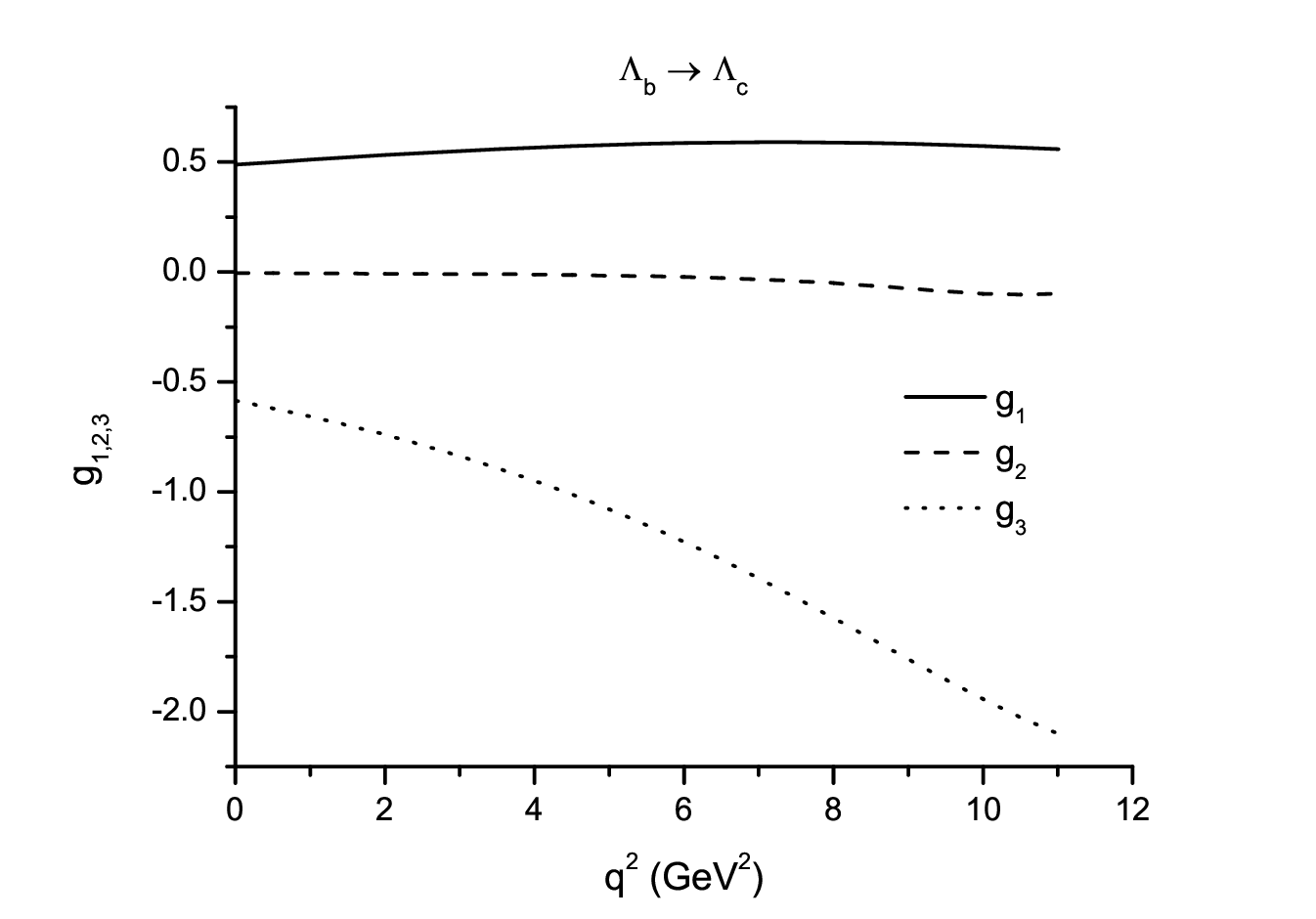}
\includegraphics[width=3.1in]{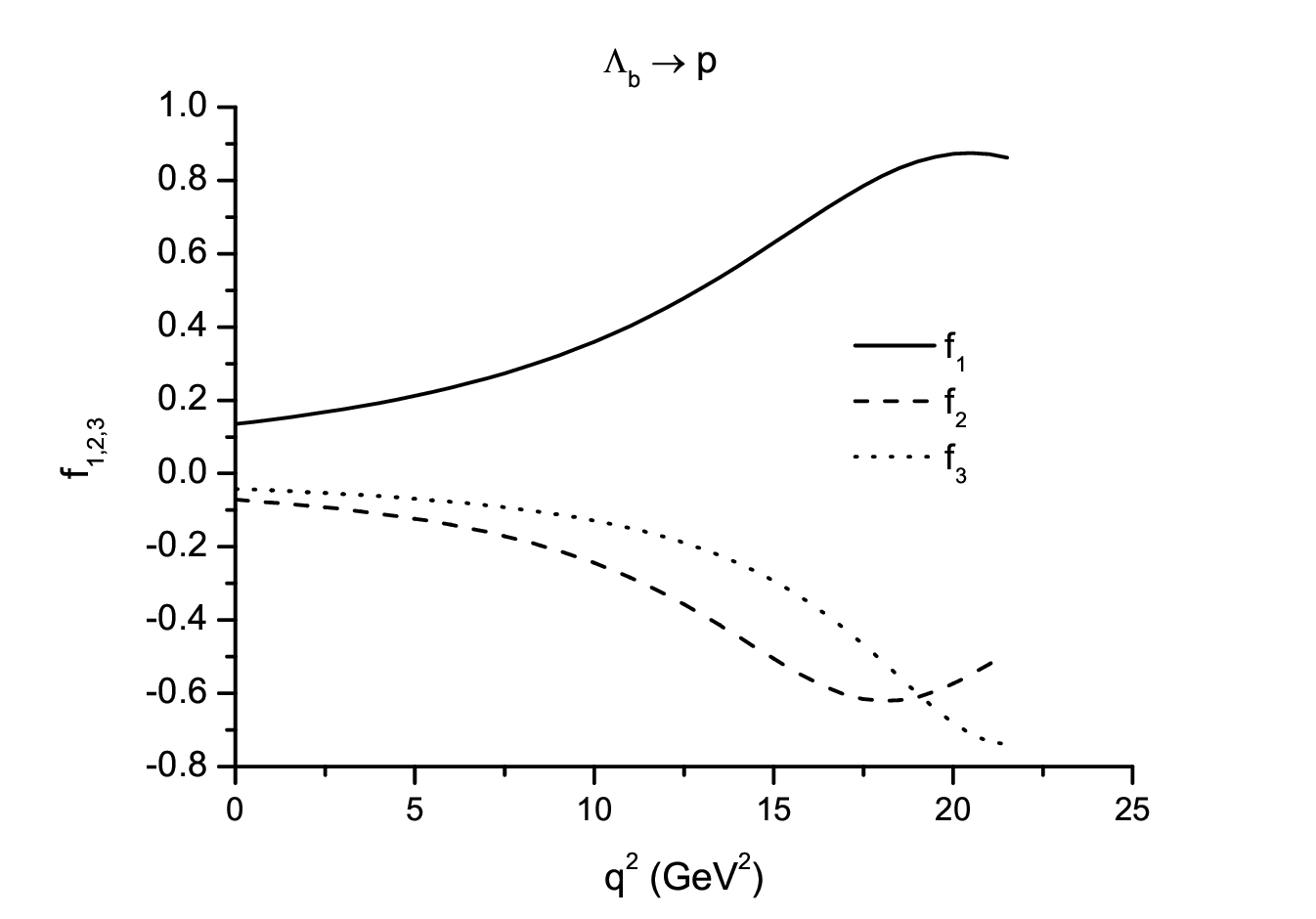}
\includegraphics[width=3.1in]{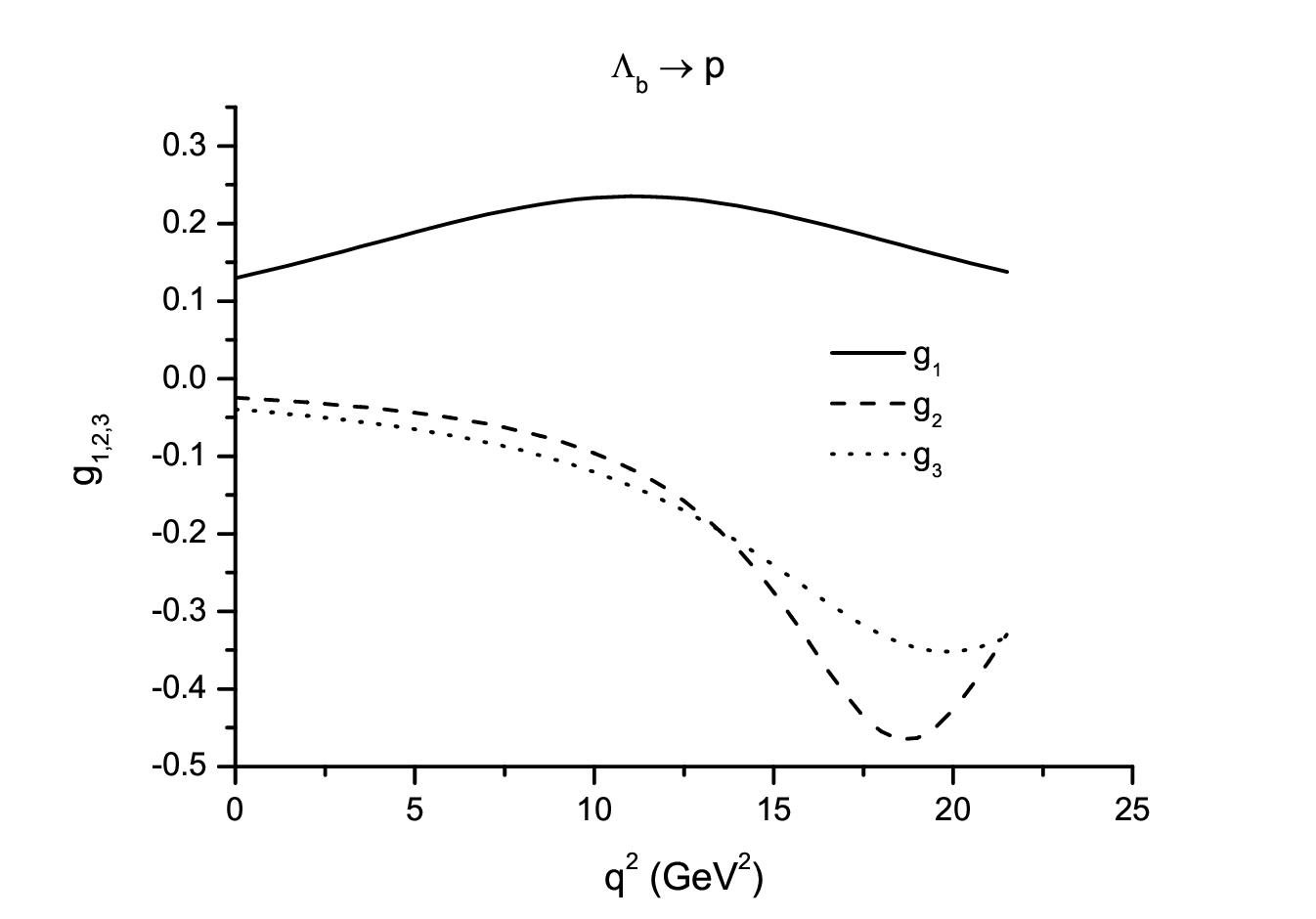}
\caption{The $\Lambda_b\to \Lambda_c$ and $\Lambda_b\to p$ transition form factors
with valence+nonvalence contributions.}
\label{fig3}
\end{figure}

In Fig.~\ref{fig3}, we present our evaluations of the form 
factors as functions of \(q^2\) for the \(\Lambda_b\to \Lambda_c(p)\) transitions.

\sse{Decay branching fractions }

The semileptonic decay amplitudes are governed by the effective weak Hamiltonian describing
the $b\to q$ ($q=u, c$) transition at quark level,
\be
M_{SL}(B_{i}\to B_{f}\,\ell \bar{\nu}_{\ell})= {G_F\over \sqrt{2}}V_{qb}
\langle B_{f}(P^{\prime},S^{\prime})|\bar{q}
\gamma^{\mu}(1-\gamma_5)b|B_{i}(P,S)\rangle
\,\bar{\ell}\gamma^{\mu}(1-\gamma_5)\nu_{\ell}\,.~
\label{he2}
\ee
The amplitude of the nonleptonic decay $B_{i}\to B_{f}+M$ is defined by
\be
M_{NL}(B_{i}\to B_{f}+M)={G_F\over \sqrt{2}}\alpha_{CKM}f_{M}M_{M}
\langle B_{f}(P^{\prime},S^{\prime})|\bar{q}
\gamma^{\mu}(1-\gamma_5)b|B_{i}(P,S)\rangle \epsilon_{\mu}(\lambda_{M})
\label{NLhe2}
\ee
where $M_{M}$ and $f_M$ represent the meson mass and decay constant 
and the subscript $M$ refers to either a pseudoscalar meson $P$ or a vector meson $V$. 
The constant $\alpha_{CKM}$ in equation ~(\ref{NLhe2}) is generally expressed as 
$\alpha_{CKM}=a_{1}V_{qb}V^{*}_{q_{i} q_{i}}$, 
where $q_{i, j}$ represent the quarks contained in the emitted meson. 
However, in the $\Lambda_{b}\to p K$ process, the main contribution comes from 
the penguin diagram; therefore $\alpha_{CKM}$ must be rewritten 
as $\alpha_{CKM}=a_{1}V_{ub}V^{*}_{us}-(a_{4}+ r_{M}a_{6})V_{tb}V^{*}_{ts}$. 
The values of $a_{4}$ and $a_{6}$ can be found in 
Refs.~\cite{CCQM,ceff}, while $r_{M}\equiv 2m^{2}_{M}/(m_b(m_{q}+m_{u}))$ 
occurs only in processes involving pseudoscalar mesons. 
The amplitudes in Eqs.~(\ref{he2}) and (\ref{NLhe2}) 
can be expressed as  combinations of independent helicity components, which are described by 
$H^{V(A)}_{\lambda, \lambda_{W}}$, where $\lambda$ and $\lambda_{W}$ 
represent the helicity components of the final hadron and the W propagator, respectively. 
The advantage is that we can easily separate the integrals for 
longitudinal and transverse polarization asymmetries. 
Using Eq.~(\ref{transitionVA}), the differential decay widths 
for semileptonic and nonleptonic processes can be expressed in terms of the 
helicity amplitudes as follows:
\be
\frac{d\Gamma_{SL}}{dq^{2}}&=&\frac{G_{F}^{2}|V_{bq}|^{2}}{(2\pi)^{3}}
\frac{(q^{2}-m_{\ell}^{2})^{2}\,\rm |p_{cm}|}{24M^{2}\,q^{2}}\nonumber \\
&\times&\bigg\{\left(1+\frac{m_{\ell}^{2}}{2 q^{2}}\right)
\sum_{\lambda_{B_{f}},\lambda'_{W}}|H_{\lambda_{B_{f}},\lambda'_{W}}|^{2}
+\frac{3 m_{\ell}^{2}}{2 q^{2}}
\sum_{\lambda_{B_{f}},t}|H_{\lambda_{B_{f},t}}|^{2}\bigg\}\,,
\label{Dwidth}
\ee
\be
\Gamma_{NL}&=&\frac{G_{F}^{2}|\alpha_{CKM}|^{2}}{32\pi M^{2}}
f_{M}^{2} M_{M}^{2}\,{\rm |p_{cm}|}
\sum_{\lambda_{\lambda_{B_{f}}}\lambda_{M}}|H_{\lambda_{2},\lambda_{M}}|^{2}\,,
\label{NLDwidth}
\ee
where $\rm p_{cm}$$ =\sqrt{Q_{+}Q_{-}}/2M$, $\lambda_{B_{f}}=\pm 1/2$, 
$\lambda'_{W}=(0, \pm 1)$ and $\lambda_{M}$ depends on the daughter meson. 
In Eq.~(\ref{NLDwidth}), we rewrite $\lambda_{W}$ in the helicity components as $\lambda_{M}$.
For the vector meson case one sums over $\lambda_{M}=(0, \pm 1)$ and for the
pseudoscalar meson case one has $\lambda_{M}=0$.

We present the calculated branching fractions for  
$\Lambda_b \to \Lambda_c(p)\,\ell\,\bar{\nu}_{\ell}~(\ell=e,\,\mu,\,\tau)$ 
semileptonic decays under the LFQM in Table \ref{Table3}, 
and compare them with the experimental data. 
In the table, for $\Lambda_b \to \Lambda_c(p)\,\ell\,\bar{\nu}_{\ell}~(\ell=e,\,\mu, \,\tau)$, our results I and II correspond to 
$G_{\mathcal B_i\mathcal B_f}=4(1)$ and 
$G_{\mathcal B_i\mathcal B_f}=0$, respectively. 
This indicates that Result II does not include nonvalence contributions. 
For ease of comparison, we also include the LFQM predictions based on the HBM 
and heavy-quark symmetry constraints from Ref.~\cite{HQS221016825}. 
From the table, whether or not the 
nonvalence contributions are included in the semileptonic decay 
$\Lambda_c(p)\,\ell\,\bar{\nu}_{\ell}$, the differences in results are negligible; 
only in the $\Lambda_b \to p$ decay does the decay rate decrease by approximately 1$\%$.

Furthermore, $f_3$ and $g_3$ appear in the temporal 
helicity amplitudes \(H_{\frac{1}{2},t}^{V/A}\) in Eq.~(\ref{Dwidth}). 
In general, since the electron and muon modes contain  small lepton masses, 
the contributions from $f_3$ and $g_3$ to the $e$ and $\mu$ modes are clearly negligible, 
whereas they can have a non-negligible effect in the $\tau$ modes. 
According to the calculation results, if $f_{3}(g_{3})$ is taken into account, 
the branching ratio decreases by approximately 10\% compared to 
when it is not included at all. 
Our results are consistent with the PDG data, except that the 
predicted branching fractions for $\Lambda_{b}\to\Lambda_{c}\,\ell\,\bar{\nu}_{\ell}$ are slightly 
smaller than the corresponding PDG values.
An interesting observable is the ratio of the tauonic to the electronic (muonic) decay rate. 
As shown in Table~III, our result agrees well with the LHCb measurement~\cite{LHCb1}.
\begin{table}[htbp]
\caption{Experimental and theoretical results on the 
branching fractions of semileptonic 
decays of $\Lambda_{b}$. 
Here, Result~I includes contributions from both valence and nonvalence states, 
while Result~II includes only the valence state contribution. In the table, $\ell = e$.}
\vskip 0.2in
\label{Table3}
\resizebox{\textwidth}{!}{%
\begin{tabular}{|c||c|c|c|c|c|c|} 
\hline

Result & ${\cal B}(\Lambda_b \to \Lambda_{c}\ell\bar{\nu}_{\ell})$ 
& ${\cal B}(\Lambda_b \to \Lambda_{c}\tau\bar{\nu}_{\tau})$ 
& $R^{\ell\tau}=\frac{{\cal B}({\Lambda_b}\to {\Lambda_c}\,\tau\,\nu_\tau)}{{\cal B}({\Lambda_b}\to {\Lambda_c}\,\ell\,\nu_l)}$ 
& ${\cal B}(\Lambda_b \to p \ell\bar{\nu}_{\ell})$ 

& ${\cal B}(\Lambda_b \to p \tau\bar{\nu}_{\tau})$ 

& $R_p^{\ell\tau}=\frac{{\cal B}(\Lambda_b\to p\tau\bar{\nu}_{\tau})}{{\cal B}(\Lambda_b\to p\ell\bar{\nu}_{\ell})}$

\\ \hline 
\hline

$\rm{I}$ 
& $(5.39^{+0.42}_{-0.47})\% $  
& $(1.41^{+0.051}_{-0.062})\% $ 
& $0.261^{+0.097}_{-0.117} $

& $(3.33^{+0.53}_{-0.55})\times 10^{-4} $ 
& $(2.08^{+0.22}_{-0.26})\times 10^{-4} $ 

& $0.624^{+0.119}_{-0.129}$

\\ \hline

$\rm {II}$ 
& $(5.40^{+0.42}_{-0.47})\% $

& $(1.41^{+0.051}_{-0.062})\% $ 
& $0.261^{+0.097}_{-0.117} $ 

& $(3.36^{+0.54}_{-0.58})\times 10^{-4} $ 
& $(2.10^{+0.22}_{-0.27})\times 10^{-4} $ 

& $0.625^{+0.119}_{-0.134}$

\\ \hline
 
Exp.\cite{PDG,DELPHI,LHCb1,LHCb2,LHCb3,CDF2,CDF3}  
& $(6.2\pm 1.4)\% $
& $(1.5\pm 0.64)\% $
& $0.242\pm 0.125 $ 
& $(4.1\pm 1.0)\times 10^{-4} $ 
& $- $
& $-$
\\ 
\hline

CCQM~\cite{CCQM1,CCQM2}
& $6.9\% $
& $2.0\% $ 
& $- $ 
& $- $ 
& $- $
& $-$
\\ 
\hline

LQCD~\cite{LQCD1,LQCD2}
& $5.32\pm0.35\% $
& $- $ 
& $- $ 
& $- $ 
& $- $
& $-$
\\ 
\hline

RQM~\cite{RQM}
& $6.48\% $
& $2.03\% $ 
& $- $ 
& $4.5 \times 10^{-4} $ 
& $2.9 \times 10^{-4} $
& $-$
\\ 
\hline

HBM~\cite{HQS221016825}
& $(5.69\pm0.58)\% $
& $(1.83\pm0.12)\% $
& $0.3243\pm0.0126$
& $- $ 
& $- $
& $-$
\\ 
\hline

LFQM~\cite{HQS221016825}
& $(5.35\pm0.50)\% $
& $(1.87\pm0.15)\% $
& $0.3506\pm0.0046$
& $- $ 
& $- $
& $-$
\\ 
\hline

LCSR~\cite{LCSR}
& $5.81^{+1.16}_{-1.21}\% $
& $1.59^{+0.28}_{-0.29}\% $ 
& $- $ 
& $- $ 
& $- $
& $-$
\\ 
\hline

QCDSR1~\cite{QCDSR1}
& $6.61\pm1.08\% $
& $- $ 
& $- $ 
& $- $ 
& $- $
& $-$
\\ 
\hline

QCDSR2~\cite{QCDSR2}
& $6.04\pm1.7\% $
& $1.87\pm0.52\% $ 
& $- $ 
& $- $ 
& $- $
& $-$
\\ 
\hline

QCDSR3~\cite{QCDSR3}
& $7.98^{+4.61}_{-2.62}\% $
& $2.31^{+1.54}_{-0.83}\% $ 
& $- $ 
& $- $ 
& $- $
& $-$
\\ 
\hline

HCQM~\cite{HCQM} 
& $6.04\% $
& $- $ 
& $- $ 
& $- $ 
& $- $
& $-$
\\ 
\hline

SM~\cite{SM} 
& $4.83\% $
& $1.63\% $ 
& $- $ 
& $3.89\times 10^{-4} $ 
& $2.75\times 10^{-4} $
& $0.7071\ (0.588\text{--}0.878)$
\\ 
\hline

PQCD\cite{PQCD,PQCD1,PQCD2}
& $2\% $
& $- $ 
& $- $ 
& $16\pm11 \times 10^{-4} $ 
& $(11\pm7) \times 10^{-4} $
& $-$
\\ 
\hline

LFQM\cite{LFQM,LFQM1}
& $(6.47\pm0.96)\% $
& $(1.97\pm0.29)\% $ 
& $- $ 
& $4.02 \times 10^{-4} $ 
& $2.74 \times 10^{-4} $
& $-$
\\ 
\hline
\end{tabular}
}
\end{table}

We also apply this method to nonleptonic decays 
\(\Lambda_b\to \Lambda_c\,\pi(K,D,D_s)\) and \(\Lambda_b\to p\,\pi(K,D_s)\), 
where each decay amplitude is decomposed into the product of a form factor and a decay constant. 
The results, which include contributions from both valence and nonvalence states, 
are shown in Tables~\ref{Table4} and \ref{Table5} together with the PDG data.
We observe that, within the range of uncertainties, our result for 
$\Lambda_b\to p\,D_{s}$ is approximately 50\% larger than that in the PDG data, 
while those for the other decay modes are consistent with each other. 
It is worth noting that for the $\Lambda_b\to p\,K$ decay branch, 
the main contribution comes from penguin diagrams, 
even though the corresponding effective Wilson coefficients are small.

\begin{table}[htbp]
\caption{Experimental and theoretical branching fractions 
for nonleptonic $\Lambda_{b}\to \Lambda_{c}$.}
\vskip 0.2in
\label{Table4}
\begin{tabular}{|c||c|c|c|c|c|} \hline
 Result & ${\cal B}(\Lambda_b \to \Lambda_{c} \pi) $ 
& ${\cal B}(\Lambda_b \to \Lambda_{c} K)$ 
& ${\cal B}(\Lambda_b \to \Lambda_{c} D)$ 
& ${\cal B}(\Lambda_b \to \Lambda_{c} D_{s})$ 
& ${\cal B}(\Lambda_b \to \Lambda_{c} D^{*}_{s})$
\\ \hline \hline
Our results & 
 $(4.29^{+0.77}_{-0.80})\times 10^{-3}$ & 
 $(3.48^{+0.61}_{-0.53})\times 10^{-4}$ 
& 
 $(5.43^{+0.61}_{-0.69})\times 10^{-4}$ & 
 $(1.21^{+0.13}_{-0.14})\%$ 
& 
 $(1.81^{+0.16}_{-0.19})\%$ 

\\ \hline
Exp.\cite{PDG,DELPHI,LHCb1,LHCb2,LHCb3,CDF2,CDF3}  
& $ (4.9\pm0.4)\times 10^{-3} $
& $ (3.56\pm0.28)\times 10^{-4} $
& $ (4.6\pm0.6)\times 10^{-4} $ 
& $ (1.10\pm0.1)\% $ 
& $ (1.66^{+0.09}_{-0.15})\% $ 
\\ \hline
CCQM~\cite{CCQM}
& $ - $
& $- $ 
& $- $ 
& $ 1.478\% $ 
& $ 2.516\% $ 
\\ \hline
BM~\cite{BagModel}
& $ (4.5\pm0.2)\times 10^{-3} $
& $ (3.4\pm0.1)\times 10^{-4} $ 
& $- $ 
& $- $ 
& $- $ 
\\ \hline
PQCD~\cite{PQCD4}
& $ (6.7^{+3.2}_{-2.2})\times 10^{-3}$
& $ (0.5^{+0.3}_{-0.2})\times 10^{-3} $ 
& $- $ 
& $- $ 
& $- $ 
\\ \hline
HQET~\cite{BSM} 
& $ (3.6\pm0.3)\times 10^{-3} $
& $- $ 
& $- $ 
& $- $ 
& $- $ 
\\ \hline
LFQM~\cite{LFQM1,LFQM2} 
& $ 4.96\times 10^{-3} $
& $ 3.93\times 10^{-4} $ 
& $ 5.22\times 10^{-4} $ 
& $ 1.24\% $ 
& $ 1.05\% $ 
\\ \hline
\end{tabular}
\end{table}

\begin{table}[htbp]
\caption{Experimental and theoretical branching fractions for nonleptonic $\Lambda_{b}\to p$.}
\vskip 0.2in
\label{Table5}
\begin{tabular}{|c||c|c|c|} \hline
 Result & ${\cal B}(\Lambda_b \to p \pi) $ 
& ${\cal B}(\Lambda_b \to p K)$ 
& ${\cal B}(\Lambda_b \to p D_{s})$ 
\\ \hline \hline
Our results 
& $ (4.35^{+2.27}_{-1.80})\times 10^{-6} $
& $  (5.26^{+2.73}_{-2.17})\times 10^{-6} $ 

& $  (1.98^{+0.86}_{-0.74})\times 10^{-5} $ 

\\ \hline
Exp.\cite{PDG,DELPHI,LHCb1,LHCb2,LHCb3,CDF2,CDF3} 
& $ (4.6\pm0.8)\times 10^{-6} $
& $ (5.5\pm1.0)\times 10^{-6} $ 
& $ (1.25\pm0.13)\times 10^{-5} $  
\\ \hline
CCQM~\cite{CCQM}
& $ - $
& $ - $ 
& $ 1.32\times 10^{-5} $ 
\\ \hline
BM~~\cite{BagModel}
& $ (5.0\pm0.5)\times 10^{-6} $
& $ (6.0\pm0.7)\times 10^{-6} $ 
& $ - $ 
\\ \hline
PQCD~\cite{PQCD2,PQCD3}
& $ (5.21^{+2.42}_{-1.8})\times 10^{-6} $
& $ (1.82^{+0.74}_{-0.71})\times 10^{-6} $ 
& $ - $ 
\\ \hline
BSM~\cite{BSM} 
& $ - $
& $ - $ 
& $ - $ 
\\ \hline
LFQM~\cite{LFQM1,LFQM2} 
& $ 4.30\times 10^{-6} $
& $ 2.17\times 10^{-6} $ 
& $ 1.61\times 10^{-5} $ 
\\ \hline
\end{tabular}
\end{table}

Finally, we discuss the sensitivities of our results to the input parameters. 
In Table~III, the central values correspond to the central values 
of \(\beta_{b[qq]}\), \(\beta_{c[qq]}\) and $\beta_{u[qq]}$ in Table~\ref{beta values}, 
while the uncertainties are obtained by taking the 
$\beta$ parameter uncertainties. 
Note that our results are based on fixed quark masses 
\(m_q\) and \(m_{[qq]}\). 
Obviously, different choices of these parameters also affect the outcome. 
For example, for \(\Lambda_b \to \Lambda_c \ell \nu_{\ell}\), 
if the diquark mass \(m_{[qq]}\) (which typically ranges from \(0.4\) to \(1.5\) GeV) 
is increased by 20\% without altering the baryonic \(\beta\) values, 
the decay branching ratio will increase as well. 
The uncertainties in this study stem solely from the uncertainties in the \(\beta\) factors.

\sse{Forward-backward asymmetry}

The forward-backward asymmetry is an important observable quantity. 
The $q^2$-dependent forward-backward asymmetry for the 
$\Lambda_b \to \Lambda_c(p)\,\ell\,\bar{\nu}_{\ell}$ semileptonic decay 
can also be expressed in terms of helicity amplitudes, and is given by
\be
A_{FB}(q^2) = \frac{\frac{d\Gamma}{dq^2}({\rm forward})-\frac{d\Gamma}{dq^2}({\rm backward})}
{\frac{d\Gamma}{dq^2}} 
= -\frac34\frac{H_{PV}(q^2)+
2\frac{m_\ell^2}{q^2}H_{T}(q^2)}
{H_{tot}(q^2)}.
\label{eq:afb}
\ee
with
\be
H_{PV}(q^2)&=&|H_{+1/2,+1}|^2-|H_{-1/2,-1}|^2\,, \nonumber \\
H_{T}(q^2)&=&{\rm Re}(H_{+1/2,0}H_{+1/2,t}^\dag+H_{-1/2,0}H_{-1/2,t}^\dag)\,, \nonumber \\
H_{tot}(q^2)&=&\left(1+\frac{m_{\ell}^{2}}{2 q^{2}}\right)
\sum_{\lambda_{B_{f}},\lambda'_{W}}|H_{\lambda_{B_{f}},\lambda'_{W}}|^{2}
+\frac{3 m_{\ell}^{2}}{2 q^{2}}
\sum_{\lambda_{B_{f}},t}|H_{\lambda_{B_{f},t}}|^{2}
\ee
The width-weighted average of the forward-backward asymmetry over the
physical region is defined as
\be
\label{afbmean}
\left\langle A_{FB}\right\rangle=-\frac34
\frac{\displaystyle\int_{m_\ell^2}^{(M_{\Lambda_b}-M_{B_{f}})^2}
\frac{(q^{2}-m_{\ell}^{2})^{2}\,\rm |p_{cm}|}{24M^{2}\,q^{2}} \left\{H_{PV}(q^2)+
2\frac{m_\ell^2}{q^2}H_{T}(q^2)\right\}\,dq^2}
{\displaystyle\int_{m_\ell^2}^{(M_{\Lambda_b}-M_{B_{f}})^2}
\frac{(q^{2}-m_{\ell}^{2})^{2}\,\rm |p_{cm}|}{24M^{2}\,q^{2}} H_{tot}\,dq^2}\,.
\ee
where $M_{B_{f}}$ and $m_{\ell}$ denote the masses of the final baryon ($\Lambda_c$ or $p$) 
and the lepton, respectively.

We list the integrated leptonic forward-backward asymmetries in
Table~\ref{pred}. The decay modes are arranged in columns, following
the format used in Table~\ref{Table3}, while the different theoretical calculations
are given in rows. For each reference, only one representative prediction
is listed. 
The entries labelled Results~I and II are obtained from the width-weighted
 integration 
in Eq.~(\ref{afbmean}); Result~I includes both the valence and nonvalence contributions, 
while Result~II includes only the valence contributions. 
The uncertainties for Results~I and II in all tables are given by 
the envelope formed by the upper and lower bounds of the $\beta$ parameters 
in the baryon wave functions listed in Table \ref{beta values}.
\begin{table*}[t]
\caption{Theoretical predictions for the integrated leptonic forward-backward
asymmetry $\langle A_{FB}\rangle$ in semileptonic $\Lambda_b$ decays.  The
entries use the forward-backward convention specified in the respective
references; consequently, a sign difference can arise when the definition of
the lepton helicity angle is reversed.}
\label{pred}
\resizebox{\textwidth}{!}{%
\begin{tabular}{|c||c|c|c|c|c|c|} \hline
$A_{FB}$
& $\Lambda_b\to\Lambda_c e\bar\nu_e$
& $\Lambda_b\to\Lambda_c\mu\bar\nu_\mu$
& $\Lambda_b\to\Lambda_c\tau\bar\nu_\tau$
& $\Lambda_b\to p e\bar\nu_e$
& $\Lambda_b\to p\mu\bar\nu_\mu$
& $\Lambda_b\to p\tau\bar\nu_\tau$\\
\hline \hline
I
& $0.230^{+0.005}_{-0.004}$
 & $0.223^{+0.005}_{-0.004}$ 
& $-0.041^{+0.004}_{-0.003}$
 
& $0.067^{+0.007}_{-0.008}$
 & $0.065\pm0.007$ 
& $-0.101^{+0.004}_{-0.002}$ \\ 
\hline
 
II 
& $0.230^{+0.005}_{-0.004}$ 
& $0.223^{+0.005}_{-0.004}$ 
& $-0.041^{+0.004}_{-0.003}$ 

& $0.065\pm0.007$
 & $0.063\pm0.007$
 & $-0.102^{+0.003}_{-0.001}$ \\ 
\hline

LCSR1~\cite{LCSR2}
& $0.18\pm0.02$ & $0.17\pm0.02$ & $-0.05\pm0.03$ & $-$ & $-$ & $-$\\ 
\hline
RQM~\cite{RQM}
& $0.195$ & $0.189$ & $-0.021$ & $0.346$ & $0.344$ & $0.185$\\ 
\hline
LFQM2~\cite{LFQM3}
& $0.18$ & $0.17$ & $-0.08$ & $-$ & $-$ & $-$\\ 
\hline
CCQM~\cite{CCQM1,CCQM2}
& $0.36$ & $-$ & $-0.077$ & $-$ & $-$ & $-$\\ 
\hline
CLFQM~\cite{LFQM1}
& $-0.03$ & $-0.07$ & $-0.13$ & $0.12$ & $0.18$ & $0.10$\\ 
\hline
LFQM~\cite{LFQM}
& $0.18\pm0.05$ & $0.17\pm0.05$ & $-0.08\pm0.03$ & $-$ & $-$ & $-$\\ 
\hline
LQCD~\cite{Rahmani2025AFB}
& $-0.206$ & $-0.211$ & $-0.334$ & $-$ & $-$ & $-$\\ 
\hline
LCSR2~\cite{LCSRp}
& $-$ & $-$ & $-$ & $0.33\pm0.01$ & $0.32\pm0.01$ & $0.15\pm0.01$\\ 
\hline
\end{tabular}
}%
\end{table*}

\se{Conclusion}

Using the light front B-S formalism in the time domain, 
we have calculated all form factors associated with the transition matrix elements 
covering the entire $\Lambda_b\to\Lambda_c(p)$ physical region.
We have distinguished between form factor sets that include and exclude nonvalence states, 
and calculated the semileptonic and nonleptonic decay branching fractions for 
$\Lambda_b \to \Lambda_c(p)$, separately. 
We have also examined the forward-backward asymmetries of the semileptonic decays. 

Our results have been presented in Tables~\ref{Table3}$-$\ref{pred}. 
It is worth noting that the HQET analysis constrained by LQCD and 
LHCb spectra gives a precise prediction for $dA_{FB}/dw$~\cite{BSM}, while
the lattice-input angular analysis of the cascade decay reports predictions in
bins separated by the zero crossing~\cite{Nandi2024AFB}.  These results are
not included as single entries in Table~\ref{pred}, because they do not quote
a full-phase-space integrated value in the same form.

The calculations with valence contributions alone and with 
both valence and nonvalence contributions yield nearly identical
branching fractions and $R^{\ell\tau}(\Lambda_c)$, showing that these
integrated rates are relatively stable under this model variation. The
nonleptonic channels are also described at the same level of accuracy as the
available data, except for  $\Lambda_{b}\to p Ds$, 
where the central prediction is about 58\% higher than the measured value, 
while the quoted ranges overlap. 
In contrast, the tau mode is more sensitive to the temporal
helicity amplitudes and therefore to $f_3$ and $g_3$ form factors.
The integrated forward-backward asymmetries are approximately lepton-universal
for the light leptons, whereas the finite tau mass produces a visible sign reversal,
especially in the $\Lambda_b\to\Lambda_c$ channel.  These results provide a
compact set of observables for testing the form factor treatment and for
future measurements of the angular distributions.

\section*{Acknowledgments}

We thank Dr. C. W. Liu for helpful discussions. 
This work is supported in part by 
the National Natural Science Foundation of China (NSFC) under Grant No. 12547104.

\end{document}